\documentclass[a4paper,12pt]{article}
\usepackage{setspace}
\usepackage[UKenglish]{babel}

\usepackage[utf8]{inputenc}
\usepackage[margin=1in]{geometry}

\usepackage{amsmath, amssymb, amsthm, mathtools}
\usepackage{tikz} 

\usepackage{url}

\usepackage[toc,page]{appendix} 
\usepackage{chngcntr}

\usepackage{graphicx}
\graphicspath{ {./} }

\usepackage[backend=bibtex,style=authoryear,natbib=true]{biblatex} 

\newcommand{\mcalK}{\mathcal{K}}
\newcommand{\matern}{{Mat\'{e}rn}}
\newcommand{\mbbE}{\mathbb{E}}
\newcommand{\mbbR}{\mathbb{R}}

\usepackage[affil-it]{authblk} 
\usepackage{etoolbox}
\usepackage{lmodern}

\usepackage{orcidlink}

\usepackage{lipsum}
\patchcmd\abstract{\small}{\footnotesize}{}{}

\providecommand{\keywords}[1]
{
  \small	
  \textbf{\textit{Keywords---}} #1
}

\title{Mat\'{e}rn Correlation: A Panoramic Primer}
\author[1]{Xiaoqing Chen \orcidlink{0000-0003-4715-7227}}

\affil[1]{The Alan Turing Institute \\ British Library, 96 Euston Rd., London NW1 2DB, London, UK}
\affil[1]{Department of Mathematics and Statistics, Faculty of Environment, Science and Economy, 
University of Exeter \\ Stocker Rd, Exeter EX4 4PY, Devon, UK}

\date{\vspace{-3ex}}

\renewenvironment{abstract}
 {\par\small\textbf{Abstract}\par\vspace{0.5em}}
 {\par\vspace{0.5em}}

\begin{document}
\maketitle
\sloppy

\begin{abstract}
\matern{} correlation is of pivotal importance in spatial statistics and machine learning. 
This paper serves as a panoramic primer for this correlation with an emphasis on the exposition of its changing behavior and smoothness properties in response to the change of its two parameters. 
Such exposition is achieved through a series of simulation studies, the use of an interactive 3D visualization applet, and a practical modeling example, all tailored for a wide-ranging statistical audience.
Meanwhile, the thorough understanding of these parameter-smoothness relationships, in turn, serves as a pragmatic guide for researchers in their real-world modeling endeavors, such as setting appropriate initial values for these parameters and parameter-fine-tuning in their Bayesian modeling practice or simulation studies involving the \matern{} correlation. 
Derived problems surrounding \matern{}, such as inconsistent parameter inference, extended forms of \matern{} and limitations of \matern{}, are also explored and surveyed to impart a panoramic view of this correlation.

\end{abstract}


\section{Introduction}
\label{intro}
\citet[p.17-18]{matern1960spatial} introduced a class of isotropic (therefore stationary) correlation functions with the form of $exp(-a^2 d^2)$, whose general form is 
\begin{equation}
    \label{generalform}
    \int_{-\infty} ^{\infty} exp(-a^2 d^2) d H(a),
\end{equation}
where $H$ is a positive finite measure. By selecting a ``type III distribution" \citep[p.~126, 249]{cramer1946mathematical} for $a^2$ above, equation \eqref{generalform} became a spectral density function of frequency $s$ in $\mbbR^n$
\begin{equation}
    \label{freq}
    \int_{-\infty} ^{\infty} exp[- a^2 (d^2 + b^2)] \frac{b^{2s}}{\Gamma(s)} a^{(2(s - 1))} d a^2 = (I + \frac{d^2}{b^2})^{-s},    
\end{equation}
provided $s > n / 2$ and with a suitable constant. 
And by applying an inverse Fourier transform to the equation \eqref{freq}, a correlation function in $\mbbR^n$ is obtained as 
\begin{equation}
    \label{matern_corr}
    \frac{2^{(1-\nu)}}{\Gamma(\nu)} (b d)^{\nu} \mcalK_{\nu} (bd), 
\end{equation}
where $b$ and $d$ are non-negative, and $\mcalK_{\nu}$ is the modified Bessel function of the second kind of order $\nu$.

\citet{stein1999interpolation} then named a class of correlation functions in the form of equation \eqref{matern_corr} as the \matern{} class. The \matern{} correlation, together with the \matern{} covariance, obtained by multiplying the \matern{} correlation by a marginal variance $\sigma^2$, is of pivotal importance
in the spatial statics and machine learning research communities. This paper mainly focuses on the \matern{} correlation and defaults the marginal variance $\sigma^2$ to unit 1 without loss of generality. 

The discussion on the \matern{} class function in \citet[p.~31-33]{stein1999interpolation} mainly focused on the abstract analysis of the parameter $\nu$ mentioning the positive relationship between $\nu$ and the smoothness of the stationary process, i.e., the larger the $\nu$, the smoother the stationary process, and emphasized mainly in the discussion about the mean square differentiability of the stationary process at the origin, 
but did not give much detail about the other parameter symboled as $b$ in equation \eqref{matern_corr}.

\citet[Section.~3]{genton2015cross} referred to this parameter $b$ as the ``large-scale range parameter" but did not provide any further elaborations.

In the machine learning research community, \citet[p.~84-85]{williams2006gaussian} 
introduced the \matern{} correlation function 
with a reparametrized formula as 
\begin{equation}
\label{mater_ML}
    Corr (d) = \frac{2^{(1-\nu)}}{\Gamma(\nu)} \left(\frac{\sqrt{2 \nu} d}{l} \right)^{\nu} \mcalK_{\nu} \left(\frac{\sqrt{2 \nu} d}{l} \right),
\end{equation}
where they call the parameter $l$ ``length-scale". \citet{williams2006gaussian} also provided 1-D plots to show how the \matern{} covariance looks like when given three different $\nu$ values as well as the sample paths drawn from the Gaussian process modeled using the \matern{} covariance corresponding to these different $\nu$s while fixing the length-scale parameter $l$ to 1.

Nevertheless, in actual modeling practice (see the real-world example in Section \ref{example}), the parameter $l$ will not permanently be fixed to 1 and will be assigned different scalars to participate in the smoothing activities as well. However, the current literature about \matern{} does not provide a comprehensive exposition on 
how the \matern{} correlation will react to the change of parameter $l$. 

More concretely, how will the smoothness property of the \matern{} correlation change, and in which way the shape of the \matern{} correlation will modify, and how will the change of both $\nu$ and $l$ change the \matern{} correlation jointly? Or, put another way, how will the change echo their names in terms of ``small-scale" and ``large-scale"? How small is small, and how large is large?

These ambiguities hinder practical modeling efforts, including setting appropriate initial values for these parameters and parameter tuning in Bayesian modeling practice and simulation studies involving the \matern{} correlation.



To this end, this paper 
looks into the corresponding changes of \matern{} correlation through a series of simulation studies to provide a detailed exposition to the above queries.

Meanwhile, the paper incorporates an interactive 3D visualization applet and a real-world modeling example to assist the tangible understanding of the behavioral changes and smoothness properties of \matern{} correlation in response to the changes of these two parameters, aimed at benefiting a wide statistical audience.

Section \ref{matern_correlation} introduces a notation of \matern{} correlation, which balances between the spatial statistics community and the machine learning community. Section \ref{empiricalana} explores the corresponding changes of three constituent parts of the \matern{} correlation function with the changes of two parameters individually, in which concepts such as the modified Bessel function will be brushed up. 
In Section \ref{joint}, on top of the previous section, we first expound the effects of two parameters on the \matern{} correlation individually and jointly and how these changes echo their names. Then, in Section \ref{app}, we briefly introduce the manual of the interactive 3D visualization applet for the broader statistical audience. In Section \ref{exp_incon}, we further explore the inconsistent parameter inference problem of \matern{} as a by-product of experiencing the applet.
Section \ref{special} includes the special cases of \matern{} correlation corresponding to different parameters $\nu$, and the paper explains this from a modified Bessel function perspective. Moreover, Section \ref{example} uses an example in actual modeling practice to reinforce the understanding of the impacts of the large-scale parameter on the smoothness of \matern{} correlation. 
The paper ends with a discussion in Section \ref{discussion} in which inconsistent parameter inference problems, different extended forms of \matern{} (such as non-stationary \matern{}, multivariate \matern{}, \matern{} in $\mbbR^3$) and limitations of \matern{} are surveyed and discussed to impart a panorama of the \matern{}.

\section{Mat\'{e}rn Correlation function}
\label{matern_correlation}
To balance between the notation used in spatial statistics and machine learning community and following the clarifications in \citet[p.~26]{banerjee2014hierarchical} about the meaning of parameters $b$ and $l$ in equations \eqref{matern_corr} and \eqref{mater_ML}, we denote the Mat\'{e}rn correlation function as
\begin{equation}
    \label{Matern1}
    Corr(d) = \frac{2^{(1 - \nu)}}{\Gamma(\nu)} (\frac{d}{\rho} )^{\nu} \mcalK_{\nu} (\frac{d}{\rho}),
\end{equation}
where
\begin{itemize}
    \item $d$: the Euclidean distance between pairs of points on $\mbbR^n$,
     \item $\mcalK_v$: the modified Bessel function of the second kind of order $\nu$;
     
     meanwhile, follow \citet[Section.~3]{genton2015cross} 
     \item $\rho$: the large-scale range parameter, $\rho > 0$, and is the "length scale" in the machine learning community, as seen in Section \ref{intro}, 
    \item $\nu$: the small-scale smoothing parameter, $\nu \geq 0$.
   
\end{itemize}

In spatial statistics, equation \eqref{Matern1} is often reparametrized as  
\begin{equation}
    \label{eq_mater2}
    Corr(d) = \frac{2^{(1 - \nu)}}{\Gamma(\nu)} (\kappa  d )^{\nu} \mcalK_{\nu} (\kappa  d),
\end{equation} where
the $\kappa$ here is the reciprocal of range $\rho$ (i.e., $\kappa = \frac{1}{\rho}$) and is called the \textit{spatial decay} parameter \citep[p.~26]{banerjee2014hierarchical}.

It is apparent that the Mat\'{e}rn correlation function consists of three parts, which are 
\begin{itemize}
    \item the constant part: $\frac{2^{(1 - \nu)}}{\Gamma(\nu)}$,
    \item the power-$\nu$ part: $(\frac{d}{\rho} )^{\nu}$ or $(\kappa  d )^{\nu}$, and 
    \item the modified Bessel function part: $\mcalK_{\nu} (\frac{d}{\rho})$ or $\mcalK_{\nu} (\kappa  d)$.
\end{itemize}

In the sequel, we will investigate each of these three parts and conduct a series of simulation studies to provide an in-depth understanding of the changing effects of each of the two parameters $\rho$ and $\nu$ on each of these constituent parts and, consequently, on the value of Mat\'{e}rn correlation as a whole.

\section{Simulation Studies of Three Parts of \matern{} Correlation}
\label{empiricalana}
\subsection{The Modified Bessel Function Part: $\mcalK_{\nu} (\frac{d}{\rho})$}

By \citet[p.~77]{watson1922treatise}, the modified Bessel function $\mcalK_{\nu}(z)$ of the second kind of order $\nu$ is the second solution to the differential equation 
\begin{equation}
\label{modi_bess}
    z^2 \frac{d^2y}{d z^2} + z \frac{dy}{dz} - (z^2 + \nu ^2) y = 0,     
\end{equation}
which differs from the ordinary Bessel function only in the coefficient of the last term, i.e., 
\begin{equation*}
    z^2 \frac{d^2y}{d z^2} + z \frac{dy}{dz} + (z^2 - \nu ^2) y = 0.     
\end{equation*}

For unrestricted $\nu$, $\mcalK_{\nu}(z)$ can be expressed as 
\begin{equation}
\label{2ndsolo}
    \mcalK_{\nu}(z) = \frac{\pi}{2} \frac{I_{-\nu}(z) - I_{\nu}(z)}{sin(\nu \pi)}, 
\end{equation}
where $I_{\nu}(z)$ is the first solution to the differential equation \eqref{modi_bess}. For more details on the first solution, readers may refer to \citet[p.~77]{watson1922treatise}. 

We want to understand the property of the
modified Bessel function $\mcalK_{\nu}$ and the changing behavior of this function $\mcalK_{\nu} (\frac{d}{\rho})$ corresponds to the change of its argument. 

We start from the simplest point. 
We generate $100$ spatial points in $\mbbR^1$ equally spaced from $0.5$ to $10$ and randomly choose a value for $\nu$ for now, for example, $\nu = 1.5$, meanwhile fixing the range parameter $\rho$ to 1 for now. 

\begin{figure}[!ht] 
    \centering
    \includegraphics[width=0.65\textwidth]{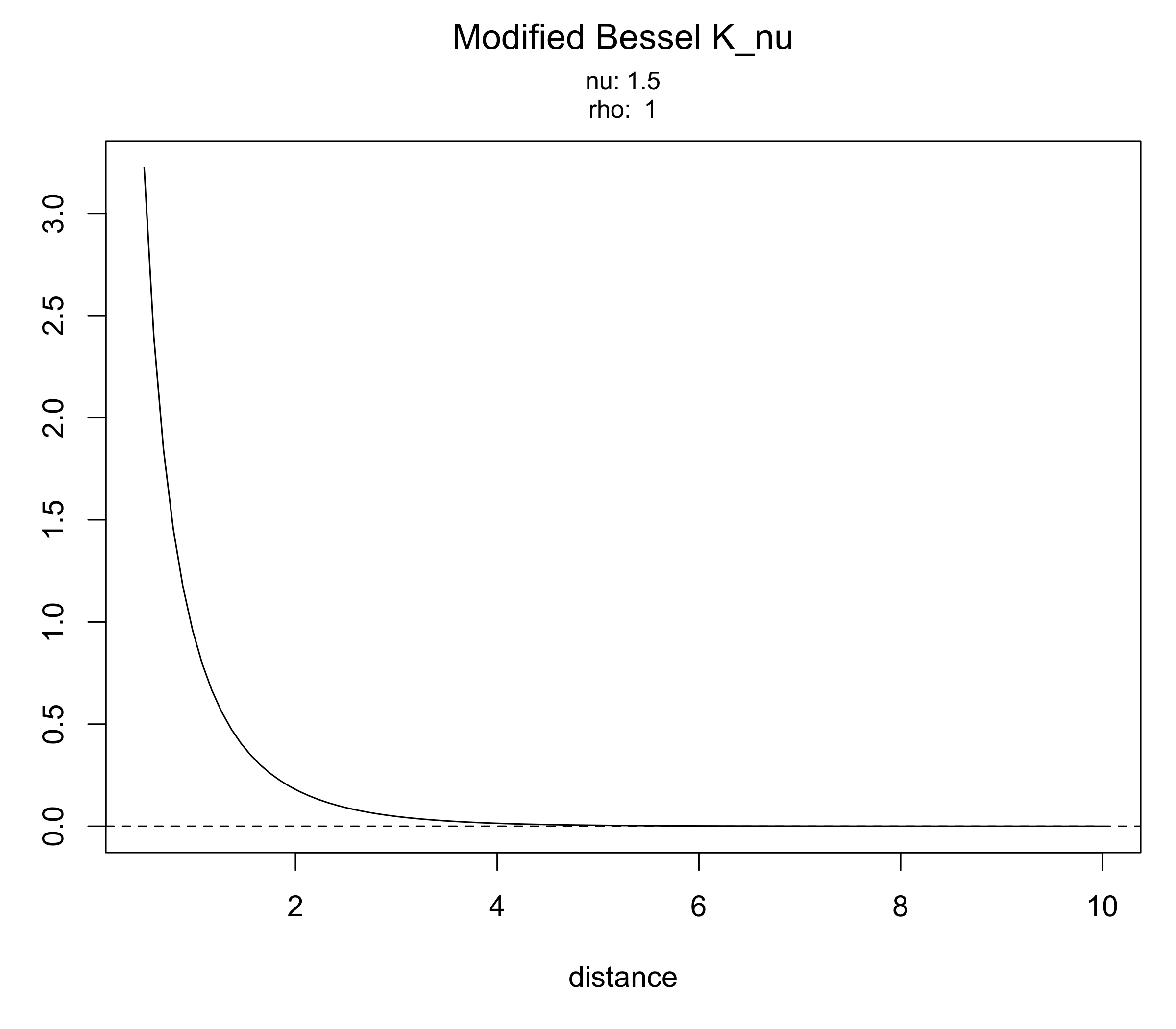}
    \caption{Modified Bessel function of the second kind of order $\nu$, $\nu$ = 1.5 and $\rho = 1$. It is an exponentially decreasing function.}
    \label{fig:bessel_nu_1.5} 
\end{figure}

From Figure \ref{fig:bessel_nu_1.5}, we understand that the modified Bessel function $\mcalK_{\nu}$ of the second kind of order $\nu$ is an \textit{exponentially decreasing} function, which agrees with the Figure 9.7 in \citet[p.~374]{abramowitz1968handbook}. 
This property alone gives us abundant information about how the value of the Bessel function part will change as the range parameter $\rho$ changes. More specially, the larger the $\rho$, the smaller the $\frac{1}{\rho}$, hence the greater the value of the Bessel function part $\mcalK_{\nu} (\frac{d}{\rho})$ for a given distance $d$ and a fixed $\nu$.

Moreover, if in a reparametrization setting where the reciprocal of the range parameter $ \rho$ is reparametrized as the spatial decay parameter $\kappa = \frac{1}{\rho}$, we know that the larger the $\kappa$, the smaller the $\rho$, the smaller the value of the modified Bessel function part $\mcalK_{\nu} (\kappa d )$.

\begin{figure}[!ht] 
    \centering
    \includegraphics[width=0.9\textwidth]{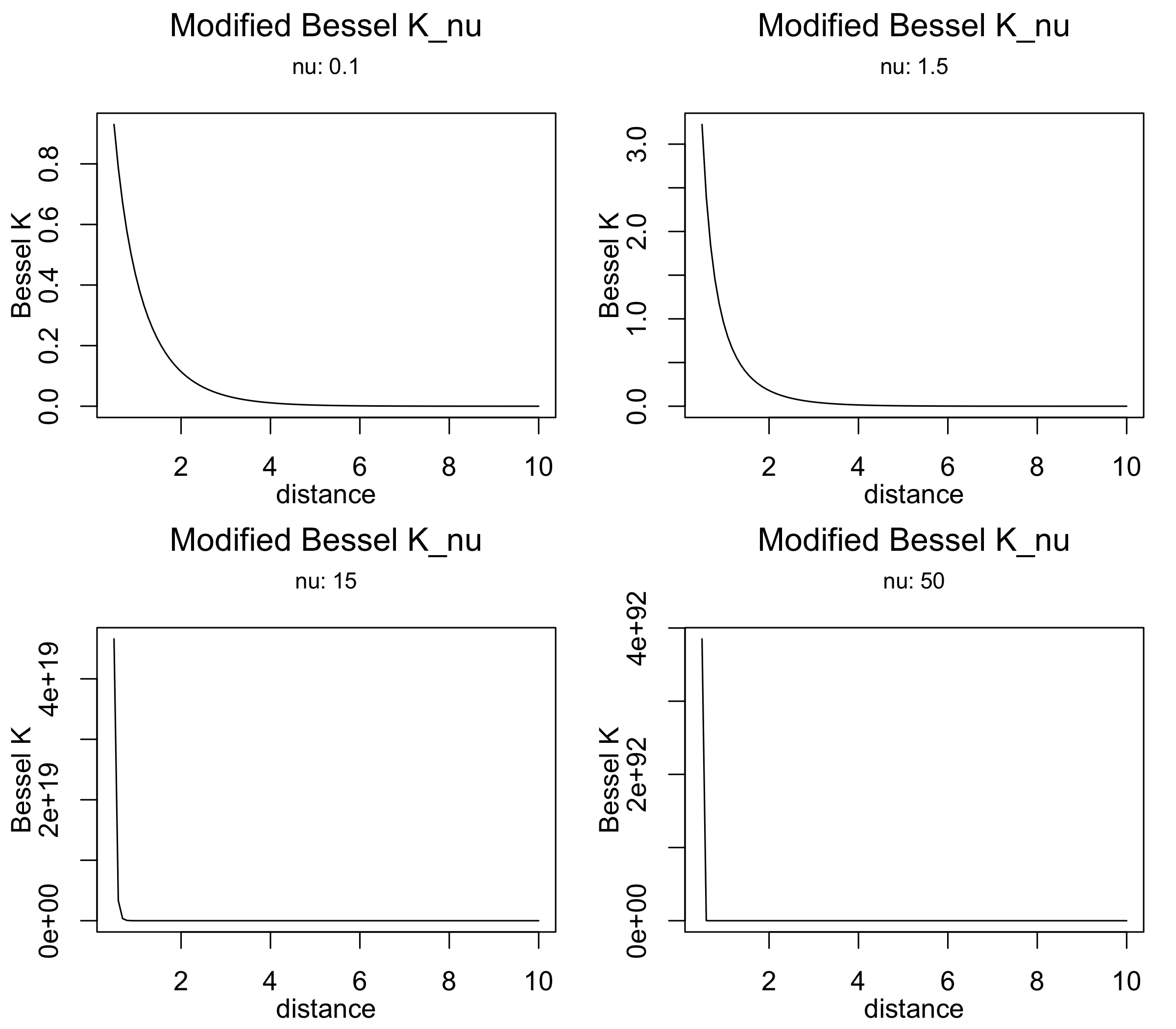}
    \caption{Modified Bessel function of the second kind of order $\nu$, with four different $\nu$s and $\rho$ is fixed to one. The value of the modified Bessel function increases dramatically with the increase of $\nu$.}
    \label{fig:bessel_diff_nu} 
\end{figure}

We then change to different $\nu$'s (but still fixing $\rho = 1$ and the same set of spatial distances) to understand the relationship between the smoothing parameter $\nu$ and the value of the modified Bessel function.
Figure \ref{fig:bessel_diff_nu} tells us that the value of the modified Bessel function increases with the increase of $\nu$. Notice that the values on the y-axis increase dramatically from 3.0 when $\nu$ is 1.5 to $4e+19$ when $\nu$ is 15. This simulation result echoes what has been emphasized in \citet[p.~31]{stein1999interpolation} and \citet[p.~84-85]{williams2006gaussian}. 

\subsection{The Power-$\nu$ Part: $(\frac{d}{\rho} )^{\nu}$}
To understand the property of the power-$\nu$ part, we randomly select some choices for both $\rho$ and $\nu$, e.g., $\rho$ ranges from $0.5$ to $50$ by 10, and $\nu$ is selected to represent two extreme scenarios and two mild ones, i.e., $0.5, 1.5, 10, 50$. 

\begin{figure}[!ht] 
    \centering
    \includegraphics[width=1\textwidth]{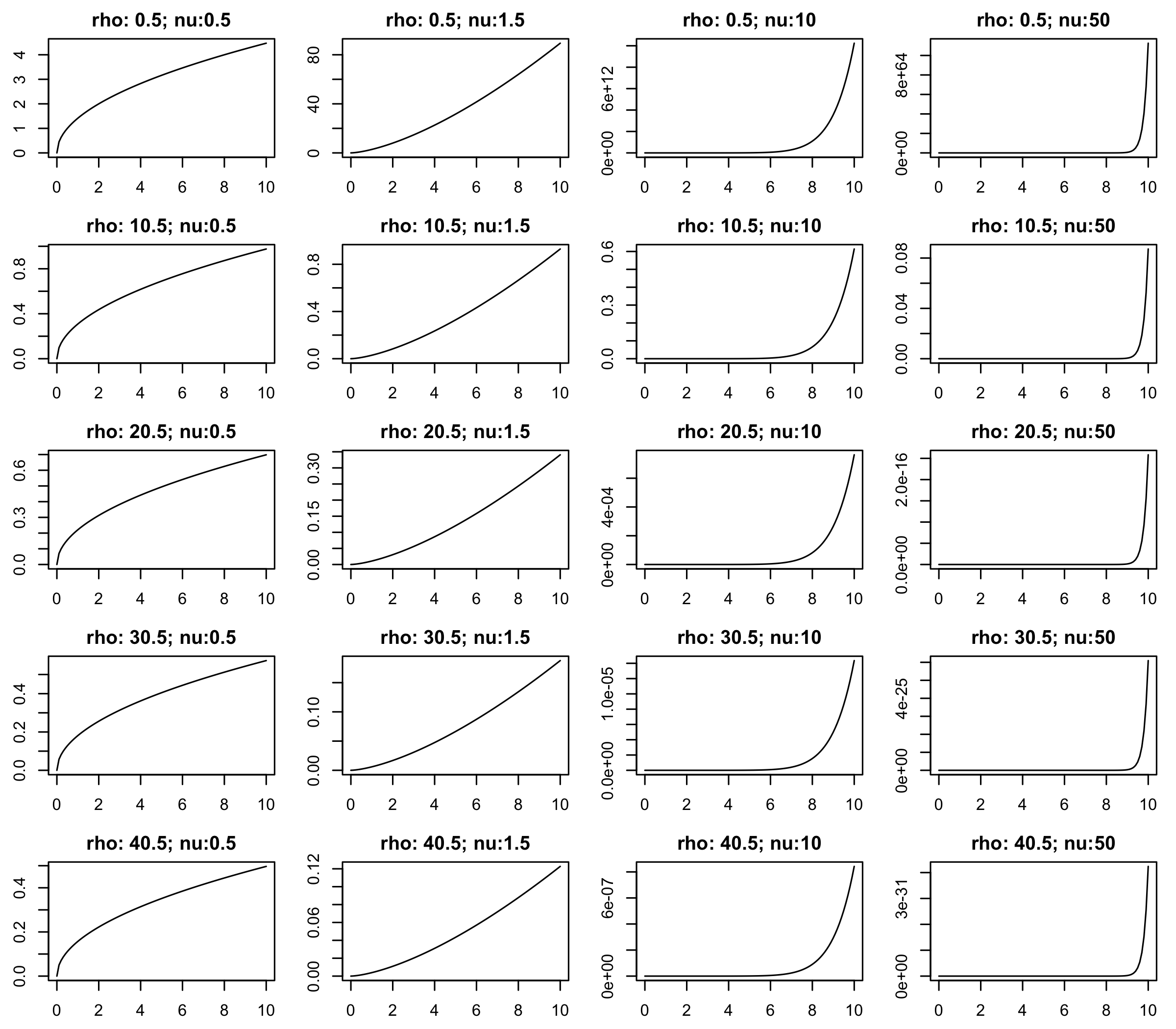}
    \caption{The Power-$\nu$ corresponds to different combinations of $\rho$ and $\nu$. The x-axis is spatial distance $d$ ranging from (0, 10) by 0.01, and the y-axis is the value of power-$\nu$. When inspecting column-wisely for a given $\nu$, changing $\rho$ does not change the shape of the power-$\nu$ curve but only scales down the magnitude of the curve; when inspecting row-wisely for a given $\rho$, increasing $\nu$ changes the shape of the curve, from logarithmic to approximately linear to exponentially increase.}
    \label{fig:power_nu_rho} 
\end{figure}

When inspecting the Figure \ref{fig:power_nu_rho} column-wisely, we discover that for a given $\nu$, changing $\rho$ does not affect the shape of the power-$\nu$ curve but scales down the magnitude of the value of power-$\nu$ curves.
Furthermore, when inspecting Figure \ref{fig:power_nu_rho} row-wisely, we notice that for a given $\rho$, when $\nu$ gets large, the power-$\nu$ curves change from logarithmic curves to approximate linear lines and then to exponentially increasing curves. 

Upon knowing that the $\rho$ has no significant effect on the shape of the power-$\nu$ part, 
we then query what $\nu$ values will induce the dramatic change in the shape of the power-$\nu$ curves from logarithmic curves to 
approximate linear lines and then exponentially increasing curves. 

\begin{figure}[!ht] 
    \centering
    \includegraphics[width=1\textwidth]{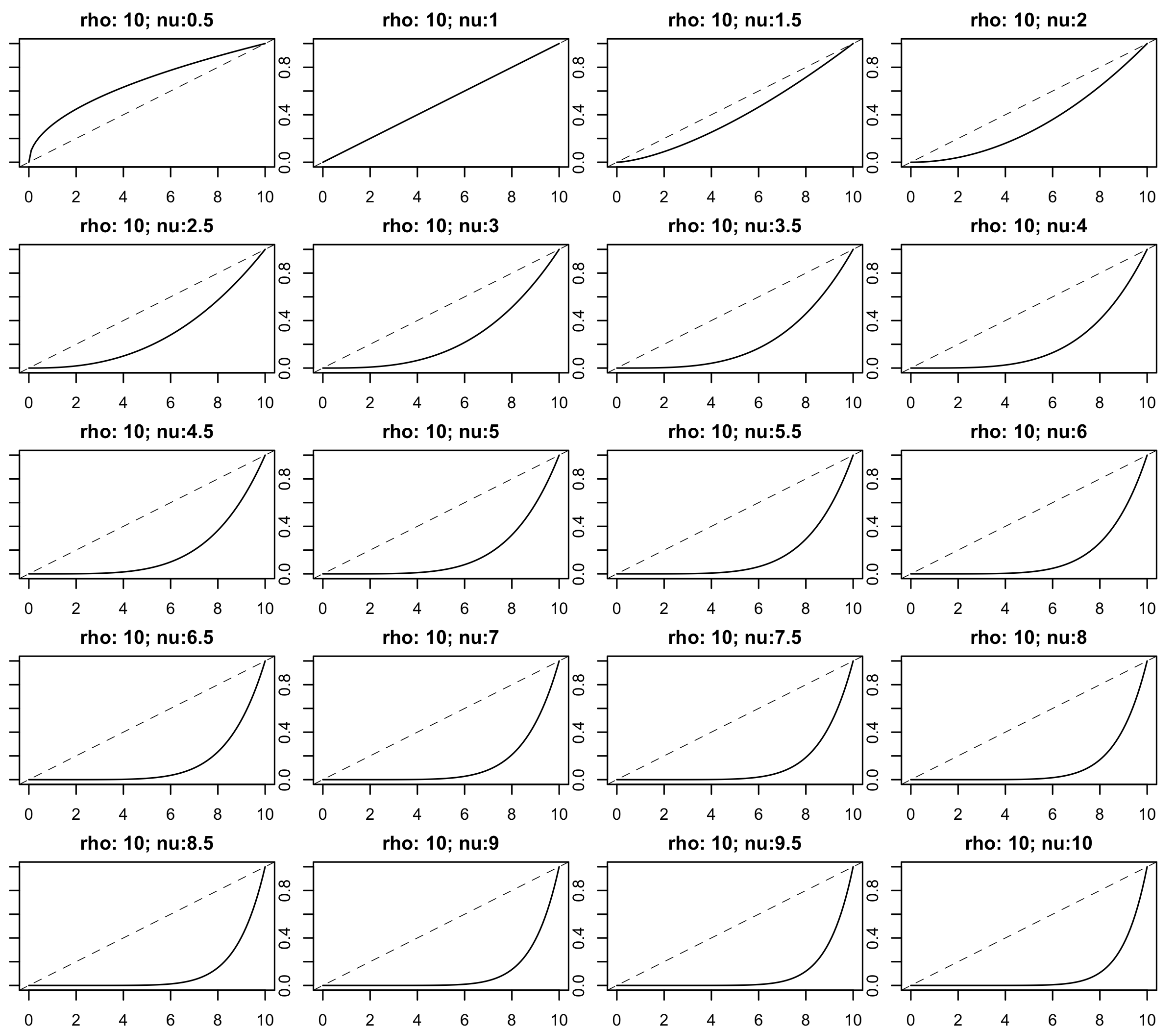}
    \caption{Power-$\nu$ curves of different $\nu$s with $\rho$ fixed to 10. The dotted line for the contrast of curvature is linear, with an intercept of 0 and a slope of 0.1. The x-axis is spatial distance $d$ ranging from (0, 10) by 0.01, and the y-axis is the value of power-$\nu$. The power-$\nu$ curve increases exponentially when $\nu$ is greater than 2.}
    \label{fig:power_nus} 
\end{figure}

We fix $\rho$ to 10 and investigate the corresponding changes of the power-$\nu$ curve when $\nu$ changes from $0.5$ to $10$. To contrast the curvature, we plot a linear line with an intercept of $0$ and a slope of $0.1$ for comparison. Figure \ref{fig:power_nus} indicates that the power-$\nu$ curve starts to become exponentially increase when $\nu$ is greater than 2.

\begin{figure}[!ht] 
    \centering
    \includegraphics[width=1\textwidth]{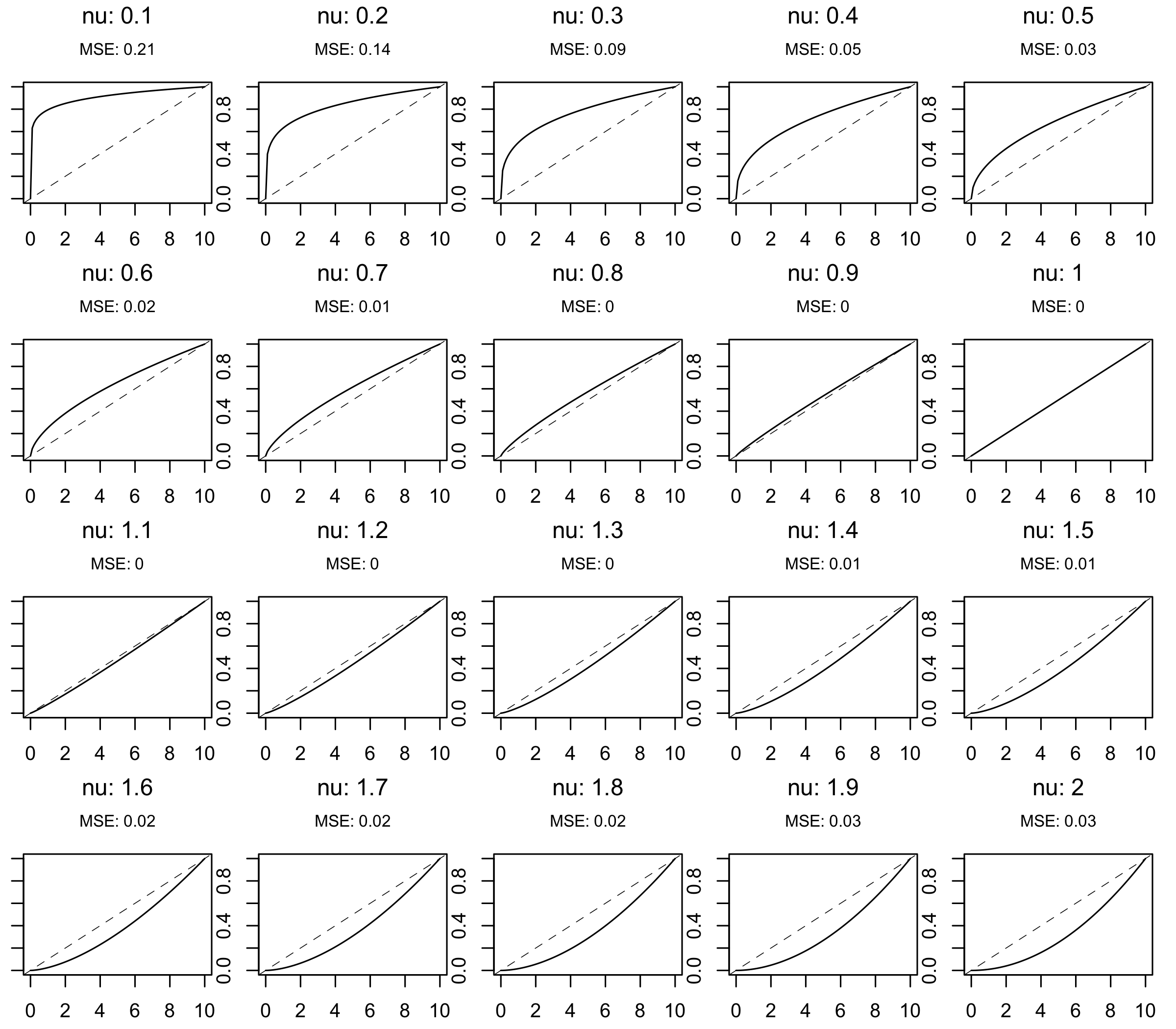}
    \caption{Power-$\nu$ curves of different $\nu$ values ranging from 0.1 to 2 with $\rho$ fixed to 10. The dotted line is linear, with an intercept of 0 and a slope of 0.1. The x-axis is spatial distance $d$ ranging from (0, 10) by 0.01, and the y-axis is the value of power-$\nu$. MSE is calculated between the power-$\nu$ curves and the linear line for each distance $d$.}
    \label{fig:power_nus_MSE} 
\end{figure}

We, therefore, narrow the scope of $\nu$ and select $\nu$ from $0.1$ to $2$ by $0.1$, with $\rho$ fixed to $10$ as well, and meanwhile, calculate the mean square error (MSE) between the power-$\nu$ curve and the linear line $y = 0.1 x$ across all 1-dimensional spatial distances. Figure \ref{fig:power_nus_MSE} tells us that when $\nu$ ranges from $0.7$ to $1.5$, the MSE between the power-$\nu$ curve and the linear line is as small as $0.01$, and outside this range of $\nu$ values, the MSE gets prominent, and the shape of the power-$\nu$ curve deviates from straight linear lines.

\subsection{The Constant Part: $\frac{2^{(1 - \nu)}}{\Gamma(\nu)}$}

\begin{figure}[!ht] 
    \centering
    \includegraphics[width=0.75\textwidth]{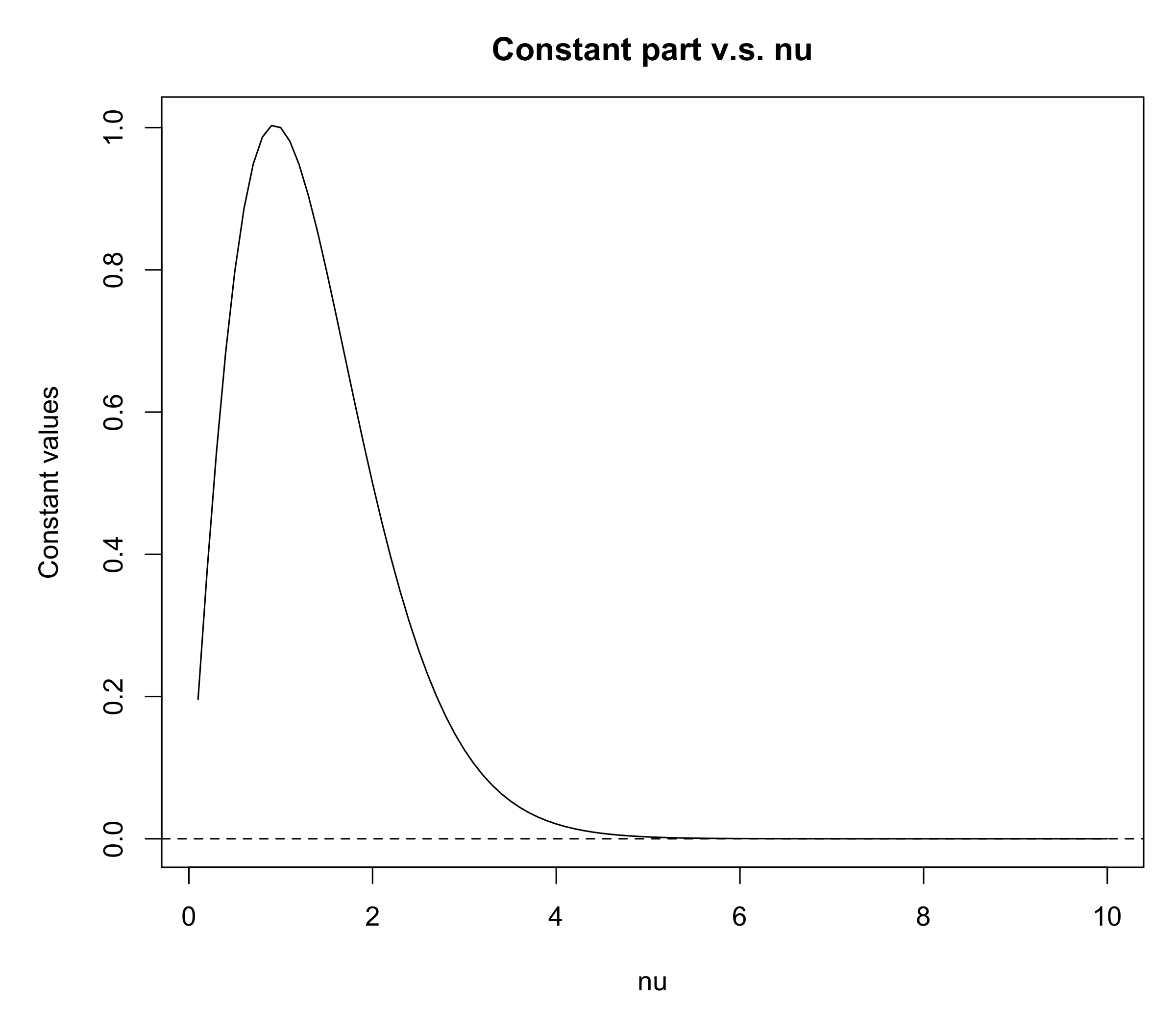}
    \caption{Values of the constant part v.s. different $\nu$ values. The value is always between 0 and 1; when $\nu$ gets greater than 5, the constant part approximates 0 infinitely.}
    \label{fig:const_nu} 
\end{figure}

For the constant part, we plot the values of $\frac{2^{(1 - \nu)}}{\Gamma(\nu)}$ corresponding to different $\nu$s against the different $\nu$ values. Figure \ref{fig:const_nu} indicates 
$\frac{2^{(1 - \nu)}}{\Gamma(\nu)}$ is a scalar always between 0 and 1, and when $\nu$ gets greater than 5, the constant part approximates 0 infinitely. 

\subsection{Summary of effects of $\nu$ and $\rho$ on Each Part of Mart\'{e}rn Function}
We give a summary of the effects of $\nu$ and $\rho$ on three different parts of Mart\'{e}rn function as follows: 

\begin{itemize}
    \item $\nu$:
        \begin{itemize}
            \item For constant part $\frac{2^{(1 - \nu)}}{\Gamma(\nu)}$: when $\nu \in (0, 5)$, its value is within $(0, 1)$; see Figure \ref{fig:const_nu};
            \item For the power-$\nu$ part $(\frac{d}{\rho} )^{\nu}$: as $\nu$ gets large, the shape of the power-$\nu$ curve changes from logarithmic to linear to exponentially increase; see Figure \ref{fig:power_nus}. Meanwhile, when $\nu \in (0.7, 1.5)$, the power-$\nu$ curve is approximately a linear line with MSE as small as 0.01, and outside this range, the power-$\nu$ curve deviates from the linear line; see Figure \ref{fig:power_nus_MSE};
            \item For the Bessel function part $\mcalK_{\nu} (\frac{d}{\rho})$: the larger the $\nu$, the larger the $\mcalK_{\nu} (\frac{d}{\rho})$; see Figure \ref{fig:bessel_diff_nu}.
        \end{itemize}

    \item $\rho$:
        \begin{itemize}
            \item For constant part $\frac{2^{(1 - \nu)}}{\Gamma(\nu)}$: null;
            \item For the power-$\nu$ part $(\frac{d}{\rho} )^{\nu}$: as $\rho$ gets large, the shape of the power-$\nu$ curve remains the same, but its magnitude is scaled down; see Figure \ref{fig:power_nu_rho} column-wisely; 
            \item For the Bessel function part $\mcalK_{\nu} (\frac{d}{\rho})$: the larger the $\rho$, the larger the value of the Bessel function; see Figure \ref{fig:bessel_nu_1.5}.
        \end{itemize}
    
\end{itemize}

\section{Parameters Effects on Mat\'{e}rn Correlation }

\subsection{Individual and Joint Parameter Effects on \matern{} Correlation}
\label{joint}
Putting the above investigation information together, we know in general, increasing $\nu$ will increase both the power-$\nu$ part and the Bessel function part, resulting in an increased 
Mat\'{e}rn correlation value; meanwhile, increasing $\rho$ will scale down the magnitude of the value of the power-$\nu$ part but will increase the value of the Bessel function. 

When $\nu$ is set within the range of 0.7 to 1.5, the power-$\nu$ part is approximately a linear line, and the constant part is between 0 and 1. Consequently, the value of the Mat\'{e}rn correlation function is proportionate to the value of the Bessel function part; that is, the larger the $\nu$, the larger the Bessel function value, the larger the Mat\'{e}rn correlation value; the larger the $\rho$, the larger the Bessel function value, the larger the Mat\'{e}rn correlation value. 

\begin{figure}[!ht] 
    \centering
    \includegraphics[width=1\textwidth]{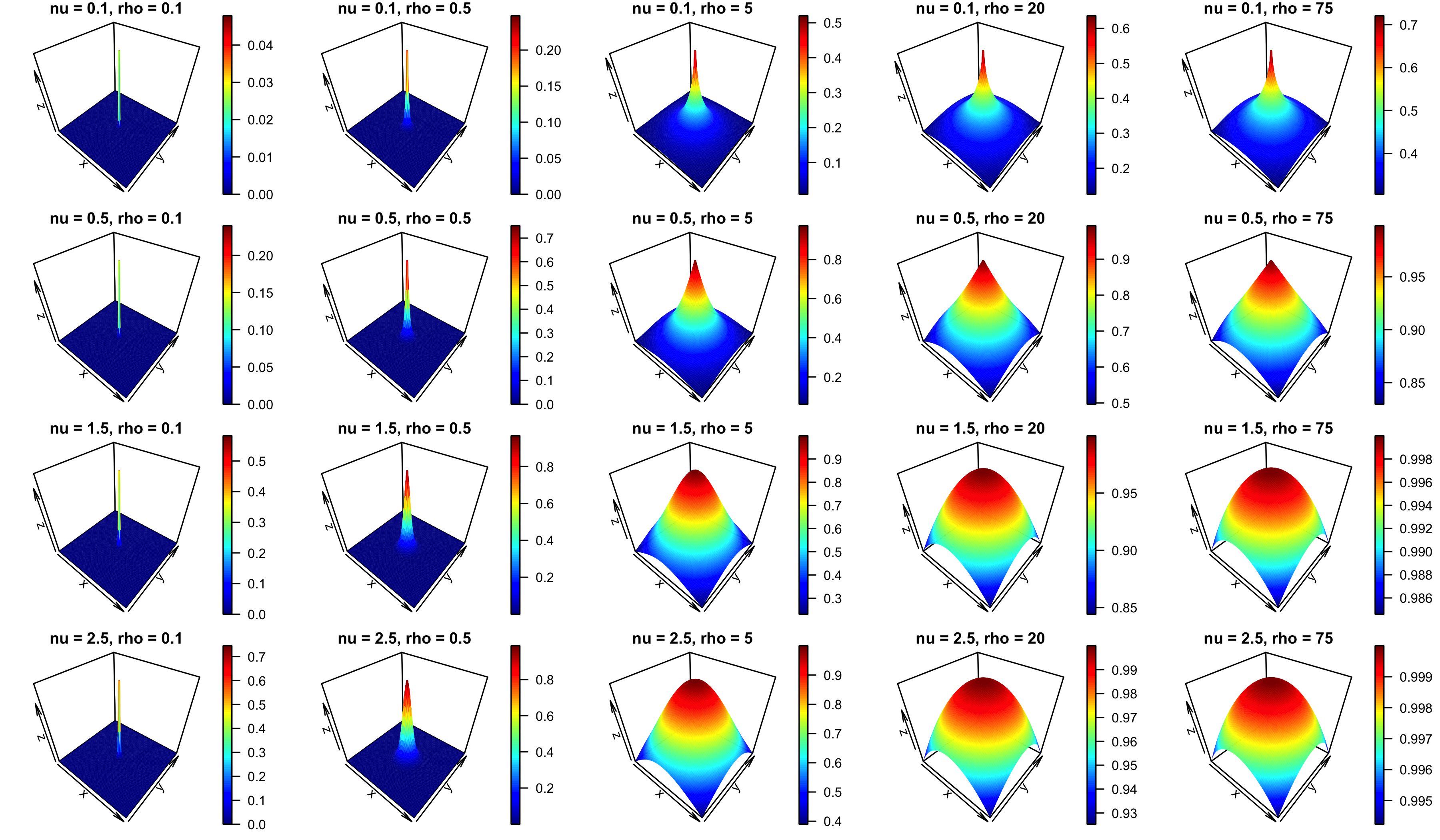}
    \caption{3D plots of Mat\'{e}rn correlation with different $\nu$'s and $\rho$'s. When observing row-wisely, the Mat\'{e}rn correlation surface spans a wider and broader domain with the increase of $\rho$, reflecting that the surface gets smoother with the increase of $\rho$ and echoes $\rho$'s name, the ``large-scale range parameter"; when observing column-wisely, for small $\rho$, only the correlation surface within a limited domain gets smoothed with the increase of $\nu$, and for large $\rho$, the value of the surface correlation becomes finer, reflecting that $\nu$ is more focused on small-scale smoothing. }
    \label{fig:Matern} 
\end{figure}

Figure \ref{fig:Matern} exhibits the 3D plots of the Mat\'{e}rn correlation function with different $\nu$s and $\rho$s. When observing row-wisely, i.e., with the increase of $\rho$ for a given $\nu$, the Mat\'{e}rn correlation surface spans a wider and broader domain. This feature reflects that, on one hand, the Mat\'{e}rn correlation changes from an instant decrease with the increase of spatial distance (e.g., when $\rho = 0.1$) to a prolonged decay with the increase of spatial distance (e.g., when $\rho = 75$), meaning the surface gets smoother with the increase of $\rho$, and on the other, echoes again the name of the parameter $\rho$, which is the ``large-scale range parameter". 

Conversely, when observing column-wisely, i.e., with the increase of $\nu$ for a given $\rho$, apart from the expected increasing value of the Mat\'{e}rn correlation, we also notice 
two things. 
One is when $\rho$ is small (for example, $\rho = 0.5 $, see the second column in Figure \ref{fig:Matern}), indicating the \matern{} correlation surface spans a small domain; 
we observe that no matter how large we increase the $\nu$ value, the surface of the \matern{} correlation only dominates a limited range, and the main change corresponding to the change of $\nu$ is the smoothness of the surface within that particular limited domain. The other thing is with the increase of $\nu$ when $\rho$ is set to an enormous value (e.g., $\rho$ = 75, see the last column in Figure  \ref{fig:Matern}), the correlation values become finer and more accurate; notice that the scale on the right-hand side of plots ranges from 0.4 to 0.7 for $\nu = 0.1$, whereas, ranges from 0.995 to 0.999 for $\nu = 2.5$. These two things reflect that $\nu$ is more focused on controlling small-scale smoothing in contrast to the range parameter $\rho$, which is in charge of large-scale smoothing. 

Moreover, when observing diagonally from top left to bottom right, with the increase of both $\nu$ and $\rho$, we notice three things. One is the increase of \matern{} correlation value as a whole; one is the domain in which the \matern{} correlation spans get broader and wider, which is the smoothing results of the increasing $\rho$ (large-scale range parameter); and the last thing is the \matern{} correlation values become finer and more accurate as shown in the colour key, which is the smoothing results of the increasing $\nu$ (small-scale smoothing parameter).

\subsection{The Shiny Applet}
\label{app}
A shiny app has been developed to assist a more tangible understanding of what has been analyzed above about the impacts of the different parameters on the \matern{} correlation.
It is available online at \url{https://xiaoqingchen.shinyapps.io/Matern_Tutorial/}. 
Readers are encouraged to experiment with various parameter combinations and observe how they influence the 3D surface of the \matern{} correlation. This will help a more comprehensive and visually vivid understanding of the parameter impacts.

In particular, set the parameter $\rho$ in the slider bar to a small number, e.g., $0.1$, and enlarge the values for the parameter $\nu$ to observe how the surface of \matern{} correlation smooth only at a small range of the domain, which echoes ``small scale" in its name,
and conversely set the parameter $\nu$ to a small quantity, e.g., $0.1$, and enlarge the values for $\rho$ to see how \matern{} correlation surface expands its smoothing domain at a much broader range, reinforcing the ``large scale" in its name. One may also try to enlarge or decrease both parameters and observe the change in the smoothness of the correlation surface. 

In the meantime, observe the changing values of the \matern{} correlation themselves in the colour key to understand the impact of the changing effects of these two parameters on the \matern{} correlation individually or jointly.

\subsection{Exploration of Inconsistent Parameter Inference Problem}
\label{exp_incon}
One interesting scenario is when we set the $\nu$ and $\rho$ to two specific values, e.g., 
$\nu = 5$ and $\rho = 40$, and then reverse the numbers assigned to these parameters, i.e., 
$\nu = 40$ and $\rho = 5$, and compare the shapes of the two corresponding \matern{} correlations under these two sets of parameters, we discover that the resulting shapes of \matern{} correlations for these two sets of parameters are very similar. 

\begin{figure}[!ht] 
    \centering
    \includegraphics[width=1\textwidth]{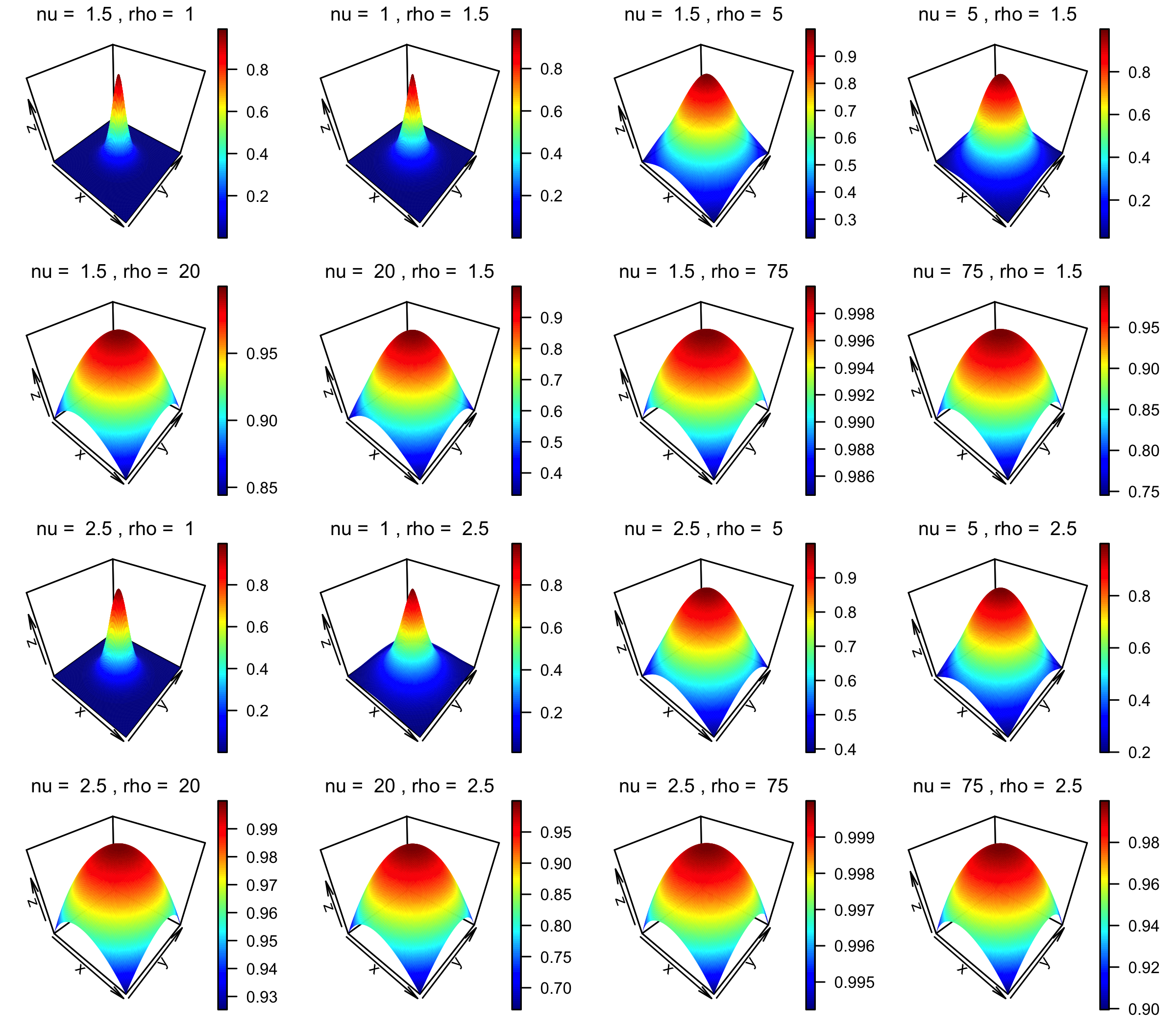}
    \caption{3D plots of Mat\'{e}rn correlation with different combinations of $\nu$'s and $\rho$'s in which the product of two parameters remains the same.}
    \label{fig:inconsist} 
\end{figure}

We then look into more combinations of $\nu$ and $\rho$. Figure \ref{fig:inconsist} shows the \matern{} correlation functions for various value combinations of $\nu$ and $\rho$ where the product of $\nu$ and $\rho$ remains the same. We compare the two correlation functions corresponding to ($\nu$, $\rho$) and ($\rho$, $\nu$). We observe that the correlation shapes of the majority pairs remain similar. 

To understand how much difference exists between two correlations for ($\nu$, $\rho$) and ($\rho$, $\nu$), we calculate the minimum and maximum value of the point-wise difference between these pairs of correlations, as shown in Table \ref{tb1}.  

\begin{table}[ht]
\label{tb1}
\centering
\begin{tabular}{rrr}
  \hline
 & Min Diff & Max Diff \\ 
  \hline
(0.1, 1) & -0.10 & 0.15 \\ 
  (0.1, 5) & -0.36 & 0.28 \\ 
  (0.1, 20) & -0.34 & 0.38 \\ 
  (0.1, 75) & -0.29 & 0.45 \\ 
  (0.5, 1) & -0.06 & 0.10 \\ 
  (0.5, 5) & -0.04 & 0.40 \\ 
  (0.5, 20) & -0.01 & 0.63 \\ 
  (0.5, 75) & -0.00 & 0.76 \\ 
  (1.5, 1) & -0.08 & 0.01 \\ 
  (1.5, 5) & 0.00 & 0.30 \\ 
  (1.5, 20) & 0.00 & 0.52 \\ 
  (1.5, 75) & 0.00 & 0.24 \\ 
  (2.5, 1) & -0.20 & 0.01 \\ 
  (2.5, 5) & 0.00 & 0.19 \\ 
  (2.5, 20) & 0.00 & 0.26 \\ 
  (2.5, 75) & 0.00 & 0.10 \\ 
   \hline
\end{tabular}
\caption{Table of point-wise minimum and maximum difference between pairs of \matern{} correlations under different combinations of $\nu$ and $\rho$ where the product of two parameters remains the same for each pair of correlations. The first column contains various choices for $(\nu, \rho)$ and $(\rho, \nu)$.} 
\end{table}

We notice that there are combinations of $\nu$ and $\rho$ whose maximum differences between the correlations corresponding to $(\nu, \rho)$ and $(\rho, \nu)$ are tiny, like $(1.5, 1)$ and $(2.5, 1)$; meanwhile, there are also combinations of $\nu$ and $\rho$ whose corresponding \matern{} correlations have a maximum difference up to $0.76$. 

Since for a mean-zero Gaussian process, its probability measure or probability density function is entirely determined by the second moment, i.e., the covariance or correlation function, then 
the above observational results indicate that if there exists a universal rule for the relationship between $\nu$ and $\rho$ to hold for different values
such that 
the \matern{} correlations corresponding to these different values of $\nu$ and $\rho$ remain equivalent, and thus the Gaussian probability densities modeled using such \matern{} correlations remain equivalent, then it would be impossible to identify and hence infer the parameter $\nu$ and $\rho$ in the likelihood uniquely and individually. Therefore, \matern{} correlation potentially has an \textit{inconsistent parameter inference problem}. 

Our conjectured rule, i.e., the product of $\nu$ and $\rho$ remaining the same for different values, only works for some value combinations under which the corresponding correlations and hence the probability densities remain almost equivalent with negligible differences, see Table \ref{tb1}. In Section \ref{discussion}, we will come back to this topic with more discussion.

\section{Special Cases}
\label{special}
Since when the order $\nu$ in the modified Bessel function is half an odd integer, i.e., $\nu = p + \frac{1}{2}$, where p is a non-negative integer, the second solution $\mcalK_{\nu}(z)$ to the modified Bessel function is simplified from equation \eqref{2ndsolo} to 
\begin{equation}
    \mcalK_{p + \frac{1}{2}}(z) = (\frac{\pi}{2z})^{\frac{1}{2}} exp^{-z} \Sigma_{r = 0}^{p} \frac{(p+r)!}{r! (p - r)! (2z)^r} 
\end{equation}
with 
\begin{equation}
    \mcalK_{\frac{1}{2}} (z) = (\frac{\pi}{2z})^{\frac{1}{2}} exp^{-z} 
\end{equation}
\citep[p.~80, eq. (12), eq. (13)]{watson1922treatise}, and as pointed out by \citet[p.~31]{stein1999interpolation}, \citet[p.~81]{williams2006gaussian} and 
\citet[p.~28-29]{banerjee2014hierarchical}, under such circumstances, one can write the Mat\'{e}rn correlation function as a product of a polynomial term of degree $p$ and an exponential term.

Specifically, when $\nu = \frac{1}{2}$ (i.e., p = 0), the Mat\'{e}rn correlation function degenerates to an exponential decay function 
$exp(- \frac{d}{\rho} )$ or equivalently $exp(- \kappa d)$. A small $\nu$ indicates a small \matern{} correlation, which drops dramatically at even a tiny spatial scale, resulting in a rough and spiky surface. 

When $\nu = \frac{3}{2} $ (i.e., p = 1), the \matern{} 
correlation function is in a closed form as $(1 + \frac{d}{\rho}) exp(-\frac{d}{\rho})$ or equivalently, 
$(1 + \kappa d) exp(- \kappa d)$; when $\nu = \frac{5}{2}$ (i.e., p = 2), the \matern{} 
correlation function is $(1 + \frac{d}{\rho} + \frac{d^2}{3 \rho^2}) exp(- \frac{d}{\rho})$.

And when $\nu$ increase to infinity, the \matern{} correlation function becomes Gaussian $exp(- \frac{d^2}{2 \rho^2})$, with an enormous \matern{} correlation value approximately to the largest correlation value of one, and even at a very long spatial distance, only very tiny changes occur in the correlation, resulting in a very smooth surface.

\section{An Example}
\label{example}
Since the positive relationship between the small-scale smoothing parameter $\nu$ and \matern{} correlation has been covered extensively in the current literature and has been validated and visualized again in this study, we emphasize solely the changing effects of the large-scale range parameter $\rho$ or equivalently, the spatial decay parameter $\kappa$, $\kappa = \frac{1}{\rho}$, in this section through an example in real-world modeling and simulation practice. 

\citet{cressie2016multivariate} proposed a variance-covariance matrix construction approach for multivariate spatial statistics via a conditional modeling scheme, specifically, 
assuming zero mean processes of two variables $Y_1$ and $Y_2$, they model as follows
\begin{align*}
     \mbbE(Y_2(s) | Y_1(\cdot)) &= \int_D b(s, v) Y_1(v) dv,  \\
     Var(Y_2(s) | Y_1(\cdot)) &= C_{2|1}(s, u), \qquad s, u \in D \subset \mbbR^d 
\end{align*}
where $b(\cdot; \cdot) : \mbbR^d \times \mbbR^d \rightarrow \mbbR^1$, and $C_{2|1}$ is the covariance of $Y_1(s)$ and $Y_1(u)$ after accounting for the conditional mean $\mbbE(Y_2(s) | Y_1(\cdot))$, hence is a univariate covariance function. 

From their derivations in Section 2, they constructed the joint variance-covariance matrix for $Y_1$ and $Y_2$ at a pair of locations $(s, u)$ block-wisely, i.e., 
\begin{itemize}
    \item $C11(s, u)$: the univariate covariance for $Y_1$ at location $s$ and $Y_1$ at location $u$, i.e., $C11(s, u) = Cov(Y_1(s), Y_1(u))$;
    \item $C12(s, u)$: the cross-covariance for $Y_1$ at location $s$ and $Y_2$ at location $u$, i.e., $C12(s, u) = Cov(Y_1(s), Y_2(u)) = \int_D C11(s, w) b(w, u) dw$;
    \item $C21(s, u)$: the cross-covariance for $Y_2$ at location $s$ and $Y_1$ at location $u$, i.e., $C21(s, u) = Cov(Y_2(s), Y_1(u)) = C12^{T}(u, s)$;
    \item $C22(s, u)$: the univariate covariance for $Y_2$ at location $s$ and $Y_2$ at location $u$, i.e., $C22(s, u) = Cov(Y_2(s), Y_2(u)) = \int \int b(s, v) C11(v, w) b(w, u) dv dw + C_{2|1}$, $s, u \in D $.
\end{itemize}
After vectorization over all locations, they arrive below formulae for constructing each block of the joint variance-covariance matrix for two processes, $Y_1$ and $Y_2$, across all locations in the domain $D$:
\begin{align*}
    Cov(Y_1, Y_1) &= C11, \\
    Cov(Y_1, Y_2) &= C12 = C11 B^T, \\ 
    Cov(Y_2, Y_1) &= C21 = B C11, \\
    Cov(Y_2, Y_2) &= C22 = B C11 B^T + C_{2|1},
\end{align*}
which must be positive definite provided $C11$ and $C_{2|1}$ are positive definite. 

Following this method, we conduct a 1-D simulation where locations $s, u \in \mbbR^1$ ranging from $[-1,1]$, and use a special form of \matern{} correlation (i.e., $\nu = \frac{3}{2}$) to construct the $C11$ and $C_{2|1}$, that is 
\begin{align*}
    C11(s, u) &= \sigma^2_{11} (1 + \kappa_{11} |u - s|) exp(- \kappa_{11} |u - s|), \\
    C_{2|1}(s, u) &= \sigma^2_{21}(1 + \kappa_{21} |u - s|) exp(- \kappa_{21} |u - s|)
\end{align*} 
where $\sigma^2_{11} = \sigma^2_{21} = 1$.
We use the same tent function form for B as theirs; details of B and the above derivations refer to Section 3.2 of their paper. 

We mainly observe the changing behavior of the smoothness of each block of the variance-covariance matrix for $Y_1$ and $Y_2$ corresponding to the change of large-scale spatial decay parameter $\kappa$ or, equivalently, the range parameter $\rho$, and $\kappa = \frac{1}{\rho}$. 

\begin{figure}[!ht] 
    \centering
    \includegraphics[width=0.7\textwidth]{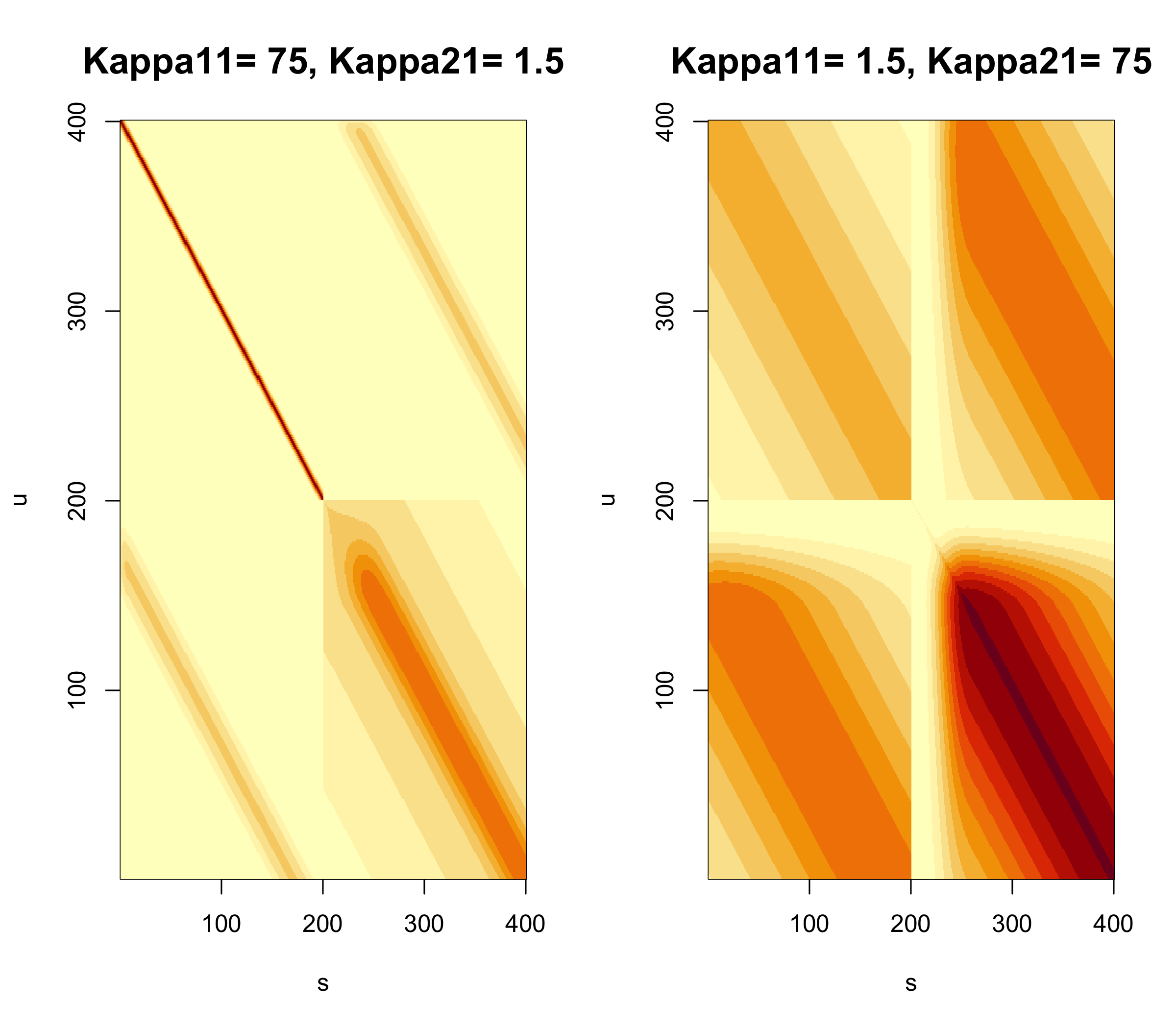}
    \caption{Comparison of different $\kappa$s and the corresponding smoothness behavior of variance-covariance matrix in actual modeling practice.}
    \label{fig:comparekappas} 
\end{figure}

Figure \ref{fig:comparekappas} exhibits the variance-covariance matrix plots for 
two different sets of $\kappa_{11}$ and $\kappa_{21}$, with the diagonal blocks representing $C11$ (upper diagonal blocks) and $C22$ (lower diagonal blocks) and the off-diagonal blocks representing $C12$ (upper off-diagonal blocks) and $C21$ (lower off-diagonal blocks).

On the left side where $\kappa_{11} = 75$ and $\kappa_{21} = 1.5$, and from this study, we know the large spatial decay parameter $\kappa$ is equivalent to a small range parameter $\rho$, meaning a small \matern{} correlation and a very rough correlation surface. Corresponding to the left figure, 
$C11$ is the block matrix on the top-left corner, and we notice the \matern{} correlation only centres on the diagonal (red line), with the rest of the area being zero (yellow colour). Moreover, due to the spike of $C11$, $C12 = C11 B^T$ and $C21 = C12^T$ are also very rough, i.e., the \matern{} correlation drops dramatically at a small spatial scale, see two off-diagonal block matrices in the left figure of Figure \ref{fig:comparekappas}. 

However, $\kappa_{21} = 1.5$ for $C_{2|1}$ is very small, indicating a very large range scale $\rho$ and \matern{} correlation value, hence a very smooth correlation surface. Furthermore, since 
$C22 = B C11 B^T + C_{2|1}$, the bottom right-hand corner block for $C22$ is a very smooth surface.

In contrast, in the right-hand figure of Figure \ref{fig:comparekappas} where $\kappa_{11} = 1.5$ and $\kappa_{21} = 75$, the top-left corner block for $C11$ now is very smooth due to $\kappa_{11} = 1.5$ as small $\kappa_{11}$ means large $\rho$ and smoother surface, meanwhile, both the off-diagonal blocks for $C12$ and $C21$ now are much smoother compared with the ones on the left-side figure. 

Furthermore, since for $C_{2|1}$, the $\kappa_{21} = 75$ is very large now, as anticipated, the $C_{2|1}$ is very small, and on the bottom-right corner block, there is only an increase and emphasis of the correlation on the diagonal area within this block, indicating very little smoothness that $C_{2|1}$ adds on top of the $B C11 B^T$, rendering the whole $C22 = B C11 B^T + C_{2|1}$, has only a little smoothness change on the diagonal region near the centre.

\section{Discussion}
\label{discussion}

This paper 
looked into 
the \matern{} correlation and particularly
the impacts of the large-scale range parameter $\rho$ as well as the small-scale smoothing parameter $\nu$ on each part of this correlation and, consequently, 
on the behavior change and the smoothness properties of this correlation as a whole.

In particular, the paper offers a comprehensive exposition of how the correlation's behavior changes in response to the change of the two parameters and how these changes echo their names in terms of ``small scale" and ``large scale". Such exposition is achieved through a series of simulation studies, the use of an interactive 3D visualization applet, and a practical modeling example, all tailored for a wide-ranging statistical audience.

Meanwhile, the thorough understanding of these relationships, in turn, serves as a pragmatic guide for researchers in their real-world modeling endeavors, such as setting appropriate initial values for these parameters and parameter-fine-tuning
in the Bayesian modeling practice or simulation studies involving \matern{}.

The series of simulation studies in this paper show increasing $\nu$ and $\rho$ will both increase the \matern{} correlation value, resulting in a slow correlation decay with increased spatial distance and a smoother surface at a small spatial scale. Conversely, decreasing $\nu$ and $\rho$ will reduce the \matern{} correlation value, resulting in a speedy correlation decay with increased spatial distance and a spiky surface, even at a small spatial scale. 

In the reparametrization setting where range parameter $\rho$ is reparametrized as spatial decay parameter $\kappa$, $\kappa = \frac{1}{\rho}$, large spatial decay parameter $\kappa$ means a small range parameter $\rho$ and will result in a small \matern{} correlation value, speedy correlation decay and very rough surface at small spatial scale, and vice versa. 

Meanwhile, $\nu$ and $\rho$ (or $\kappa$) have their own functional emphasis in terms of smoothing with 
$\nu$ is mainly in charge of small-scale smoothing, while $\rho$ is more focused on large-scale smoothing; see the detailed exposition in Section \ref{joint} or try out using the  applet introduced in Section \ref{app}.

Section \ref{exp_incon} explored the inference problem of parameters of \matern{} correlation that may be unidentifiable and inconsistent. 
\citet[p.~177]{stein1999interpolation} mentioned this problem in terms of the \matern{} covariance (i.e., \matern{} correlation multiplied by a marginal variance $\sigma^2$) in that two parameters $\sigma^2$ and $\kappa$ cannot be estimated consistently based on observations within a fixed domain $\mathcal{D}$, where $\mathcal{D} \subset \mbbR^d$, and $d = 1, 2, 3$, 
due to the existence of equivalence of random fields under a sufficient but not necessary condition, see the equation $(20)$ in \citet[p.~120]{stein1999interpolation}. 

\citet{zhang2004inconsistent} followed the idea of equivalence of Gaussian measures and the sufficient condition in \citet[p.~120]{stein1999interpolation} and showed that two probability measures are indistinguishable if 
the sufficient condition $\sigma_1^2 (\kappa_1)^{2 \nu}$ $= \sigma^2 (\kappa_2)^{ 2\nu}$ is satisfied. 
And further showed that either $\sigma^2$ or $\kappa$ will converge to different ``true" parameters under the same probability measure, hence the inconsistency in the parameter estimation. His solution is to fix the parameter $\nu$ and infer the product of $\sigma^2$ and $\kappa$, i.e., $\sigma^2 \kappa^{2\nu}$, which can be estimated consistently as shown by his Theorem 3. However, this product quantity with $\kappa$ to the power of $2\nu$ exists only in the spectral density of \matern{} covariance rather than the \matern{} covariance itself, and it is the latter one that is more frequently used in actual modeling practice.

The further reparametrization of the spatial decay parameter $\kappa$ recommended by \citet{handcock1994approach} provides a direct solution to this problem and results in a 
better-behaved model fitting \citep[p.~29]{banerjee2014hierarchical}, that is, $\kappa$ (i.e., $1/\rho$) is further reparametrized as $\alpha = 2 \sqrt{\nu} \kappa$, resulting in equation \eqref{eq_mater2} being modified as 
\begin{equation}
    Corr(d) = \frac{2^{(1-\nu)}}{\Gamma(\nu)} (2\sqrt{\nu} \kappa d)^{\nu} \mcalK_{\nu} (2\sqrt{\nu} \kappa d)
\end{equation}

or equivalently with equation \eqref{Matern1} being further modified as
\begin{equation}
    \label{eq_4}
    Corr(d) = \frac{2^{(1-\nu)}}{\Gamma(\nu)} \left(\frac{2\sqrt{\nu} d}{ \rho} \right)^{\nu} 
    \mcalK_{\nu} \left(\frac{2\sqrt{\nu} d}{\rho} \right),    
\end{equation}
which is more frequently encountered in the machine learning literature, identical to the one in equation \eqref{mater_ML} as in \citet[p.~81]{williams2006gaussian}.

One thing to note is, as mentioned at the beginning of this paper, \citet{matern1960spatial} developed this correlation function from an isotropic class of correlation; hence \matern{} correlation function is mainly used for stationary stochastic process modeling. And if more realistic phenomena are to be modeled, \citet[p.~8]{matern1960spatial} pointed out that some inhomogeneous long-distance components, either stochastic or deterministic, need to be superimposed onto the model. 

Alternatively, a non-stationary \matern{} correlation (covariance) structure has also been developed in which both the small-scale smooth parameter $\nu$ and large-scale range parameter $\rho$ become spatially varying, that is, at different locations, $\nu$ and $\rho$ are different values rather than two fixed scalars, see \cite{stein2005nonstationary}, and \cite{paciorek2006spatial}. The problem easy to foresee is the number of parameters involved and, therefore, the difficulties in the inference part.  

Another extension for univariate \matern{} is the \matern{} correlation function being used in a multivariate modeling setting where apart from using \matern{} correlation function to model the marginal correlation for each constituent process, the cross-correlations between pairs of constituent processes are also modeled using \matern{} and have their own smoothing parameter $\nu_{lk}$ and range parameter $\rho_{lk}$ where $l$, $k$ denotes different constituent processes; for more details, see \citet{gneiting2010matern}. 

\matern{} correlation does not only limit to planes in $\mbbR^2$ but can also be extended to spheres in $\mbbR^3$, where the distance metric between two locations on the surface of the sphere is typically replaced by the great circle distance in place of Euclidean distance. However, with such a great circle distance metric, \matern{} covariance on a sphere would only remain positive definite when $\nu \leq 1/2$ \citep{guinness2016isotropic}, hence the flexibility of the choice of $\nu$ is hindered, and valid \matern{} covariance on a sphere with a flexible choice of $\nu$ is much pursued. 

\citet{gneiting2013strictly} showed that the isotropic positive definite functions on $\mbbR^3$ 
allow for the direct substitution between the Euclidean distance and the great circle distance on a sphere. And based on this result, \citet{guinness2016isotropic} directly used the Euclidean distance (instead of great circle distance) between pairs of locations on a sphere, which is called \textit{chordal distance}, and developed a chordal \matern{} covariance function whose full range of choices of $\nu$ is preserved.

It is also worth noting that \matern{} correlation has only non-negative values, so such a limitation may prevent it from being a good choice for modeling the correlation between observations from both sides of a mountain ridge where \textit{orographic effects} exist \citep{guttorp1994approach}, or modeling the cross-correlation between pairs of different processes in a multivariate setting where two constituent processes are negatively or partially negatively correlated in certain regions.

\section*{Acknowledgements}
This work was supported by The Alan Turing Institute through a Turing Doctoral Scholarship. 
The author is grateful to Prof. James V. Zidek for his valuable comments and helpful suggestions that improved the quality of the paper in many aspects.



\clearpage
\printbibliography[title={References}]
\clearpage

\end{document}